\documentclass[twocolumn,preprintnumbers,amsmath,amssymb,longbibliography]{revtex4-1}

\usepackage{epsfig}
\usepackage{color}
\usepackage{amsmath}
\usepackage{multirow}
\usepackage{float}

\usepackage{siunitx}

\usepackage{lipsum}
\usepackage{graphicx}
\usepackage{comment}
\begin{document}



\title{Structure and dynamics of the magnetite(001)/water interface from molecular dynamics simulations based on a neural network potential}

\author{Salvatore Romano$^1$, Pablo Montero de Hijes$^1$, Matthias Meier$^1$, Georg Kresse$^1$, Cesare Franchini$^{1,2}$, Christoph Dellago$^1$}
\affiliation{$^1$Faculty of Physics and Center for Computational Materials Science, University of Vienna, Vienna, Austria.}
\affiliation{$^2$Department of Physics and Astronomy "Augusto Righi", Alma Mater Studiorum - Universit\`a di Bologna, Bologna, 40127 Italy}

\begin{abstract}

The magnetite/water interface is commonly found in nature and plays a crucial role in various  technological applications. However, our understanding of its 
structural and dynamical properties at the molecular scale remains still limited. In this study, we develop an efficient Behler-Parrinello neural network potential (NNP) for the magnetite/water system, paying particular attention to the accurate generation of reference data with density functional theory. Using this NNP, we performed extensive molecular dynamics simulations of the magnetite (001) surface across a wide range of water coverages, from the single molecule to bulk water. Our simulations revealed several new ground states of low coverage water on the Subsurface Cation Vacancy (SCV) model and yielded a density profile of water at the surface that exhibits marked layering. By calculating mean square displacements, we obtained quantitative information on the diffusion of water molecules on the SCV for different coverages, revealing significant anisotropy. Additionally, our simulations provided qualitative insights into the dissociation mechanisms of water molecules at the surface.

\end{abstract}

\maketitle


\section{Introduction}

Iron oxide interfaces are common in geochemistry and biology and they play crucial roles in phenomena such as corrosion and nucleation~\citep{frankel_magnetite_1979, bjorneholm_water_2016, munoz_preparation_2015, weiss_surface_2002, dictor_fischer-tropsch_1986, kirschivnik_magnetite_nodate}. Various technological areas rely on iron oxide interfaces, particularly in applications related to energy conversion and storage, nanoparticles, spintronics and even biomedicine~\citep{tartaj_iron_2011, sivula_solar_2011, wu_recent_2015, gupta_synthesis_2005, yanase_band_1984}. The importance of these interfaces and the need for a deeper understanding have driven much experimental and computational research on this topic (see~\citep{parkinson_iron_2016} for an extensive review on iron oxides). The interface between magnetite ($\rm Fe_3O_4$) and water is a prominent example of iron oxide interfaces~\citep{zaki_water_2018, kraushofer_self-limited_2019, ryan_quantitative_2022, liu_multiscale_2021, liu_bulk-terminated_2018, weiss_surface_2002, li_adsorption_2016,omranpour2024perspective}. The bare (001) and (111) surfaces of magnetite have been found to be the most stable~\citep{kovacs_electronic_2010, santos-carballal_dft_2014} and, more recently, Bliem et al.~\citep{bliem2014subsurface} discovered that the (001) surface undergoes a reconstruction known as the Subsurface Cation Vacancy (SCV) model (see Fig.\ref{fig:top_scv}), which involves the three outermost layers of magnetite. Later, water agglomeration on the SCV has been studied experimentally and computationally via Density Functional Theory (DFT) by Meier et al.~\citep{meier2018water}. Their study showed that energetically favorable configurations for various water coverages on the SCV (up to 8 $\rm H_2O$/u.c.) are stabilized by the presence of a partially dissociated water. We label as ``dissociated'' the OH group formed by dissociated water absorbed on a lattice oxygen and we refer to the ``dissociation site'' the active site where it takes place (see Fig. \ref{fig:top_scv}).

While such DFT calculations can provide insights on atomistic structures at the interface~\citep{hamada_interaction_2010, carrasco_role_2013, carrasco_wet_2011, tonigold_dispersive_2012}, they are computationally too expensive to capture spatially correlated thermal fluctuations and dynamics occurring on extended time scales. In contrast, classical force fields are computationally inexpensive but struggle to capture the complexity of the different chemical environments present at the interface~\citep{vega_simulating_2011}. In recent years, Machine Learning (ML) potentials have been developed and applied to many systems involving interfaces~\citep{montero_de_hijes_kinetics_2023, natarajan_neural_2016, quaranta_structure_2019, quaranta_maximally_2018, hellstrom_one-dimensional_2019}. A well-trained and validated ML potential can achieve the accuracy of the training dataset at a computational cost many orders of magnitude lower than that of traditional DFT.

In this study, we have developed a Neural Network Potential (NNP) following the methodology proposed by Behler and Parrinello~\citep{behler2007generalized}, enabling efficient simulations with {\em ab initio} accuracy of the interface between the SCV of magnetite and water. For a comprehensive overview on NNPs we refer to \citep{behler2015constructing, behler_four_2021}. The reference data used to train the potential were obtained from first-principle calculations based on DFT following a multi-step procedure, which started with DFT-MD simulations enhanced by kernel-regression via ``on-the-fly'' learning \citep{jinnouchi_--fly_2019, jinnouchi2019phase}. A committee of NNPs \citep{schran2020committee} trained on data generated at this stage was then used to explore the configurational space, iteratively expanding the data set until a satisfactory committee agreement was obtained. In this procedure, special care was taken to address convergence problems of the DFT calculations related to the magnetism of $\rm Fe_3O_4$. To deal with these complications, we have followed a protocol that ensures that such ``corrupted'' data are not part of the reference data set. 

We employed the newly developed NNP to carry out extensive molecular dynamics simulations of water molecules interacting with the SCV, from low coverages to liquid water. First of all we addressed the dynamics of the stable water trimers (linear and triangular shape) that are energetically equivalent at DFT level~\citep{meier2018water}. The linear trimer, which experimental observations identified as the predominant configuration, was found to be more than twice as likely to occur as the triangular trimer. The NNP proved to be an effective tool for systematic exploration of energy minima and our simulations revealed several new ground states of low water coverages on the SCV. Additionally, the calculation of mean square displacements (MSD) provided quantitative data on the diffusion dynamics of water molecules, revealing significant anisotropy. For the liquid water case, the density profile of water at the magnetite interface exhibited distinct layering, with bulk-like behavior emerging only beyond a distance of approximately 15~\AA. 

The remainder of this article is organized as follows. In Sec. \ref{sec:methods} we line out the methodology used in our work, paying particular attention to the data acquisition and curation. Results obtained for various water coverages are presented in Sec. \ref{sec:results}. A discussion and conclusions are provided in Sec. \ref{sec:discussion}.

\section{Methods}
\label{sec:methods}

In this section, we review the methodology used in our work. In particular, we describe how the dataset for the training of the NNP was constructed and cleaned of DFT inconsistencies. 
  
\subsection{Data acquisition}

\subparagraph{Preliminary dataset}
In the initial phase of data acquisition, DFT-MD simulations were carried out with the Vienna {\em ab initio} Simulation Package (VASP)~\citep{kresse_efficient_1996} using the projector-augmented wave (PAW) formalism~\citep{blochl_projector_1994}. Following Meier et al.~\citep{meier2018water}, we chose the vdW density functional optB88-DF+U, with the effective on-site Coulomb repulsion term U$_{\rm eff}$ = 3.61 eV added to describe the metal oxide system. The PAW potentials H$_{\rm GW}$, O$_{\rm GW}$, and Fe$_{\rm pv}$ were used. Based on the non-local van der Waals density functional of Dion et al.~\citep{dion2004van}, the optB88-vdW functional was proposed in Ref.~\citep{klimevs2009chemical} for the description of hydrogen bonded complexes and it has been successfully applied before in single point calculations of the magnetite-water interface~\citep{meier2018water}. The plane wave energy cutoff was set to 500 eV and we used a Monkhorst-Pack mesh to sample the Brillouin zone with (2 2 1) subdivisions along the reciprocal lattice vectors. The electronic self-consistency loop was stopped when the change in energy was smaller than  10$^{-5}$ eV. The initial guess of the local magnetization of cations was +4 $\mu_B$ for all octahedral irons and $-4$ $\mu_B$ for tetrahedral irons. In the bulk, the minimization led to some octahedral irons having magnetizations between +3.6 $\mu_B$ and +3.8 $\mu_B$, corresponding to a 2+ oxidation state, whereas the magnetizations of the other irons remained close to these initial values. 

At this initial stage, to improve configurational sampling we employed the ``on-the-fly'' learning protocol with a kernel-based model implemented in VASP~\citep{jinnouchi_--fly_2019}. For the machine learning potential, a radial cutoff of 8~\AA and 5~\AA for the two-body and three-body terms, respectively, and 8 basis functions were employed for the pair descriptors and for the three-body descriptors. The maximum angular quantum number was set to $L_{\rm max} = 4$. To speed up sampling, a large timestep  of 1.5 fs was chosen, which we compensated with an increased mass  of 8 amu for the hydrogen atoms.

The investigated magnetite surface corresponds to the SCV model in the [001] direction, which has been suggested to be highly stable ~\citep{bliem2014subsurface}. The building block of our initial configurations was a magnetite slab with 64 oxygen and 44 iron atoms exposing two SCV surface unit cells (hereafter we write u.c. to refer to the surface unit cell) due to periodic boundary conditions applied in all directions. At each of the two SCVs, water was arranged both randomly (with varying coverages from 0 to 12 $\rm H_2O$/u.c.) and as the minimum energy configurations proposed by Meier et al.~\citep{meier2018water}. From these setups, temperature ramps from 160~K up to 350~K were performed via the DFT-MD accelerated by on-the-fly machine learning. This protocol produced our first reference data made of approximately 2000 independent structures harvested from trajectories with a total length of 55 ps. With this dataset we trained a first committee of NNPs, which was then used in the second data acquisition stage.

\subparagraph{Committee of NNPs}

At this stage, we ran MD simulations with the committee of NNPs. Structures exhibiting large force disagreements are retained, then calculated {\em ab initio}, and added to the dataset, which is used to train the next generation of the committee of NNPs. These steps constitute one ``committee iteration''. The protocol converges when no more committee disagreements are found, signaling that sufficient configurational space has been covered in the dataset. 

To train and run the committee of NNPs we used n2p2~\citep{singraber_library-based_2019} interfaced with LAMMPS~\citep{singraber_parallel_2019, kyvala_diffusion_2023, thompson2022lammps}. We adopted the exponential atomic centered symmetry functions~\citep{behler_atom-centered_2011} already developed for bulk water in~\citep{morawietz_how_2016} and expanded it to account for the bulk magnetite and the interfacial regions \citep{behler2020}. We used a cutoff function of the cubic hyperbolic tangent form, which aids in preventing discontinuities in the forces~\citep{behler2015constructing} and the cutoff radius was set to 6.350 \AA. The NNP parameters were optimized with the multistream Kalman filter to minimize the cost function, which was constructed from the mean square deviation of the predicted energies and forces from their reference values~\citep{singraber_library-based_2019}.

The first generations of NNPs were frequently unstable, leading us to start with a small committee size  ($N=3$) and a relatively high disagreement threshold (0.1 eV/Å for forces). As the iterations progressed and the NNPs improved, we expanded the committee size to $N=8$ members and lowered the threshold to 0.03 eV/Å to more finely cover the configurational space. We systematically incorporated additional levels of water coverage into the dataset, ultimately extending to a liquid slab comprising 68 water molecules in contact with the magnetite slab (forming two interfaces with the PBC). Bulk water structures with 128 molecules were also included. 

In order to enable simulations of spontaneous water dissociation, an expected phenomenon on the magnetite surface, we ran simulations at higher temperatures (up to \SI{500}{\kelvin}). At these temperatures the dissociation rate was higher, providing more statistics on transition state configurations. Notably, this yielded structures with larger forces, which also contributed to the overall stability of the NNP.
Eventually, we supplemented the dataset with 342 first principle calculations of ionized water structures (involving an OH$^-$--$\rm H_3O^+$ pair) to train the committee of NNPs such that it could handle scenarios involving OH$^-$ also in water-like environments.

During the progress of the committee iterations, as the dataset increased, different architectures have been tested and the symmetry functions refined by adding new ones and pruning after sensitivity analysis. In total, more than 15 rounds of committee runs have been performed, resulting in the generation of 26,789 structures with {\em ab initio} energies and forces that constitute the final training dataset. The validation set, containing 4,807 structures with respective energies and forces, has been progressively constructed by randomly selecting structures from all performed simulations. 

As mentioned in the introduction, due the DFT complications in covering the proper spin states of the magnetite, erroneous data were systematically introduced and the standard committee protocol described here needed to be revised to include a cleaning step, which will be discussed in Sec. \ref{sec:clean}.

\subparagraph{The cleaned NNPs}

\begin{figure}[t]
    \centering
    \includegraphics[width=\linewidth]{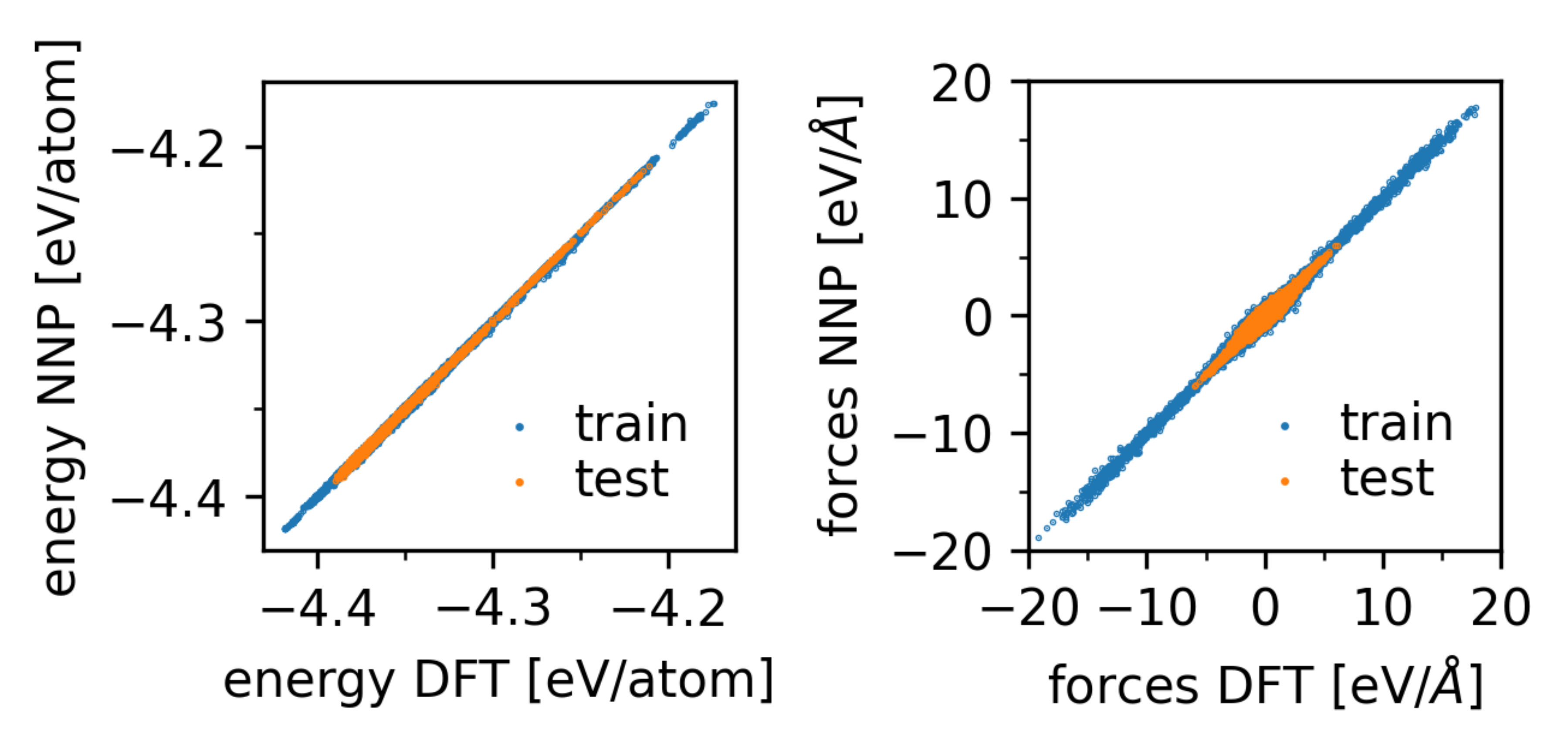}
    \caption{NNP predictions against the DFT reference for energies (left) and forces (right) for configurations in the training set (blue) and test set (orange).}
    \label{fig:prediction}
\end{figure}

The final committee comprises two distinct yet topological equivalent NNPs: one is meant to lead the production simulations while the other serves to validate the leader's predictions. The set of symmetry function contains 156 radial and angular exponential symmetry function and the architecture is as follows: 2 hidden layers with 25 nodes per layer and a hyperbolic tangent as activation function. The NNP predictions have a root mean squared error of 0.94 meV/atom in the energy and 110 meV/Å in the forces for the training set. This error is of the same order of magnitude as other Behler-Parrinello NNP not involving magnetism~\citep{montero_de_hijes_kinetics_2023, natarajan_neural_2016, quaranta_structure_2019, quaranta_maximally_2018, hellstrom_one-dimensional_2019}. On the validation set, we obtained better errors, namely 0.91 meV/atom for the energy and 100 meV/Å for the forces. The correlations of the leader NNP prediction and DFT reference for energy and forces on the training and test set are shown in Fig.~\ref{fig:prediction}.

 \subsection{Data curation}
 \label{sec:clean}
{\em Ab initio} calculations of magnetic systems based on DFT are notoriously difficult and they often fail to convergence or converge to high energy magnetic configurations~\citep{meredig_method_2010}. In our calculations of the magnetite/water interface we encountered several instances where, even after energy convergence, i) the prediction of the magnetization was incorrect, with local magnetic moments seemingly randomly assigned, and ii) convergence was to the correct magnetic state but with incorrect forces. While the exact cause of this instability in the calculation remains unclear, it may be due to the inherent complexity of the electronic energy in the presence of magnetism, where multiple local minima can be reached in the DFT energy self-consistency loop~\citep{meredig_method_2010}. 

\subparagraph{Standard data cleaning}
If these types of errors occur during {\em ab initio} MD runs, they pose a significant challenge as they would compromise subsequent time steps and thereby invalidate the rest of the simulation. Moreover, if learning is done on the fly, also the ML force field is affected because corrupted structures and energies are introduced into the dataset. To avoid these issues during the production of our preliminary dataset, we devised a procedure to continuously monitor the total magnetization and temperature. Structures exhibiting incorrect total magnetization ( accepted error 0.01 ) or too high temperature (above 1200K) were identified and removed from the training dataset. If such a configuration was detected, the simulation was interrupted and restarted from the last valid configuration with velocities redrawn from a Maxwell-Boltzmann distribution. Moreover, the kernel-regression potential of the on-the-fly learning approach was passed on to the next MD attempt after being cleaned of the last corrupted data. Although this approach is impractical for conducting long {\em ab initio} MD simulations, it nonetheless proved effective in gathering sufficient data for training the first committee of NNPs.

\subparagraph{Committee of NNPs based data cleaning}
During the committee iterations, the DFT instability was less consequential as energy calculations were performed only for single structures. Hence, the DFT calculations could be checked individually by inspecting the magnetization and the forces. Even after this ``manual'' cleaning procedure, discernible outliers were present in the force prediction vs. reference plot. Remarkably, we observed that the force outliers 1) belonged only to a small number of structures (order of 1) and 2) were outliers for all the NNP members. Therefore, only through the prediction of the committee of NNPs, we discovered the rare and most subtle type of DFT convergence issue: calculations which were converged in energy and magnetization with forces in the correct order of magnitude but appearing as outliers in the force prediction of the NNPs. A proper discrimination of these errors can invoke the simultaneous prediction of some ML-models. In order to handle all possible cases of corrupted data, we have added a cleaning step based on the predictions of the NNPs after the manual preliminary data inspection. Specifically, if all the NNPs of the committee ``agreed to disagree'' on a particular structure, this was taken as an indication of corrupted data and such structures were removed from the training dataset. Eventually, the re-trained committee of NNPs correctly predicted energies and forces of all structures in the cleaned dataset. 

Although time consuming due to manual intervention, this systematic data cleaning procedure enabled the training of an ML force field with which MD simulations could be carried out.  These simulations would have been impossible with direct {\em ab initio} MD because of the instability of the electronic structure calculation.  


\subsection{Simulations details and analysis}
\label{sec:nnp simulations deteails}
\subparagraph{MD details}
The common setup of all the initial structures is a magnetite slab exposing two independent SCV. Periodic boundary conditions are applied in all directions: $x$ and $y$ with the periodicity of the SCV unit cell and z with a period always bigger then the interaction range of the two interfaces). The thickness of the slab varies from containing 44 Fe to 50 Fe (2 bulk layers) and the water coverages from one water molecule to the full column of water (thousands of water molecules). In our MD simulations the time step was set to 0.5 fs and configurations were stored every 1000 steps. The temperature was kept constant at 200K or 300K, controlled with a Langevin thermostat applied only to four inner layers in the bulk of the magnetite slab. In this way, molecules at the interface are thermalized via interactions with surrounding molecules without any direct correction of their velocities, in particular not during dissociation and recombination events. In principle, rescaling the velocities of the water molecules could have affected diffusion, dissociation, and other physical properties of water at the interface. The committee of NNPs led the MD simulations with a threshold of the committee NNP disagreement set to 2 eV/\AA. The systems was equilibrated first at 0 bar using an NPzT simulation, in which the simulation box was free to adjust in $z$-direction. Then the volume was kept constant and the pressure fluctuated around 0 bar with a large variance.


\subparagraph{Density profiles and octahedral-tetrahedral split}
We computed the surface density profile of oxygen and hydrogen atoms of water on the SCV. As a reference point we used average $z$-coordinate, determined at each time step, of the iron and oxygen atoms in the outer surface layer, where the $z$-axis pointed normal to the surface. The distances of the water atoms from the surface were then binned into histograms. To get the density profiles, the histograms were normalized by dividing by the bin length, the number of time steps and number of unit cells so that the integral over a specific interval of distances from the surface gave the average number of atoms per unit cell that were in that interval of distances.

\begin{figure}[]
    \centering
    \includegraphics[width=\linewidth]{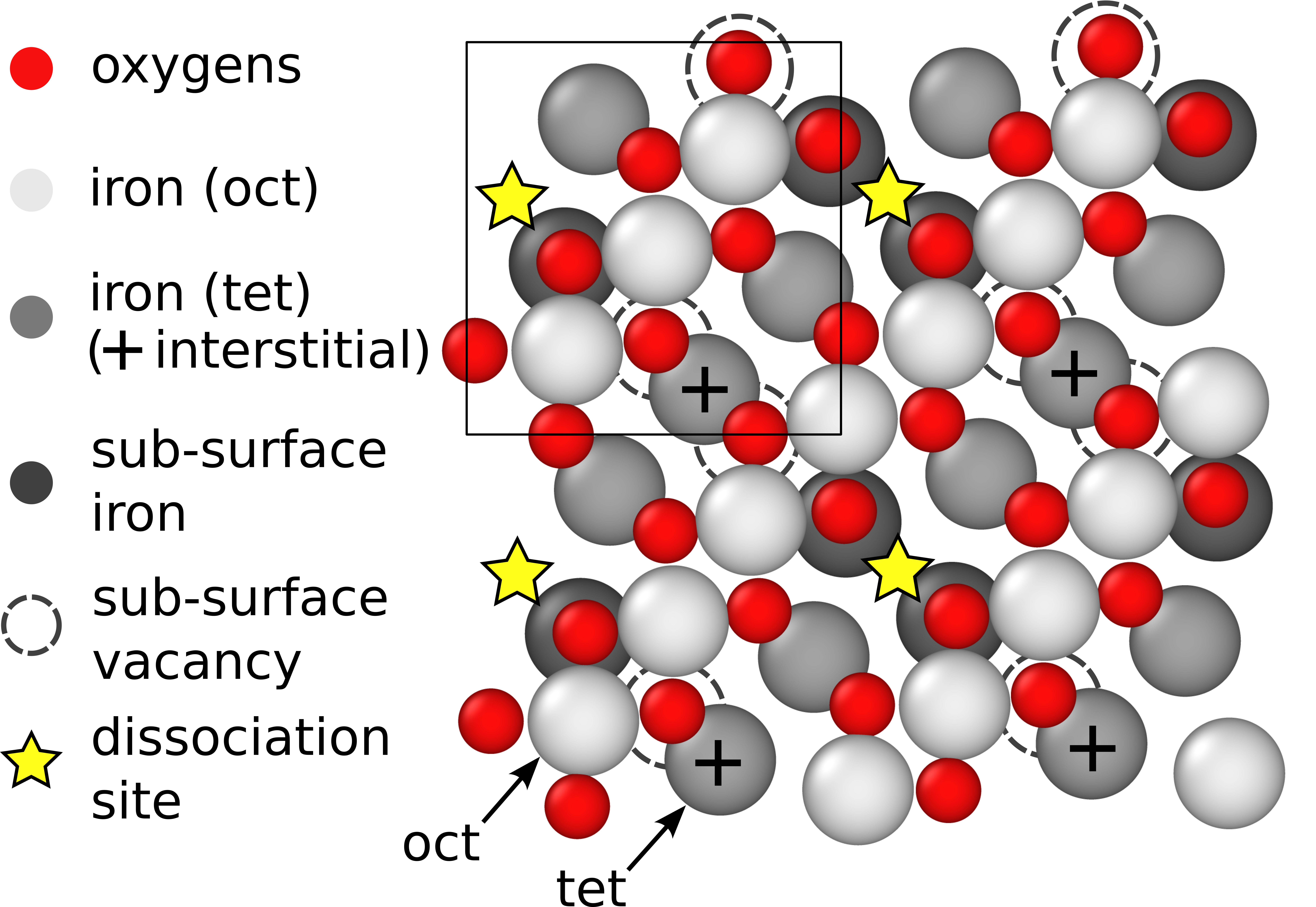}
    \caption{Top view of the SCV model. The surface unit cell ($\sqrt{2}\times\sqrt{2}$ rotated $45^\circ$ with respect to the bulk) is enclosed in the square and two times replicated. The oxygen atoms are shown in red and the iron atoms in different shades of grey to distinguish between octahedral and tetrahedral irons (light and middle gray) and the subsurface octahedral irons (dark grey). The plus symbols indicate the interstitial tetrahedral irons that during the reconstruction are lifted from the subsurface layer, leaving there the subsurface vacancies (dashed circles). The active site (stars) where the water dissociates one hydrogen (``dissociation site'') are located in the tetrahedral iron rows in the site free of irons.}
    \label{fig:top_scv}
\end{figure}

To refine the information provided by the density profiles, geometrical details of the surface were introduced. The SCV features an alternating pattern of an outer row of octahedral irons and oxygens (more precisely, irons that would have been octahedrally coordinated in the bulk) and an underlying row of tetrahedral irons, with one dissociation site per unit cell (see Fig.~\ref{fig:top_scv}). We exploited the chemical differences between these rows and subdivided the atoms above into two groups by adding the spatial information of the $xy$-projection of hydrogens and oxygens. The iron rows were determined at every timestep by grouping the surface oxygens belonging to the same row and averaging their $xy$-projections on the (anti)-diagonal (orthogonal to the rows). The latter values serve to locate the water hydrogens and oxygens on the surface: if the $xy$ projection of an atom was in between the oxygen diagonals next to the the octahedrally coordinated irons, the atom was labeled as \emph{octahedral} or \emph{oct}; otherwise, it was labeled as \emph{tetrahedral} or \emph{tet}.

\section{Results}
\label{sec:results}

In this section, we present our findings on the magnetite-water interface, beginning with the analysis of low water coverage, followed by a discussion of liquid water. 

\subsection{Low water coverage}
 
 \subparagraph{Trimer dynamics}

\begin{figure}[]
    \centering
    \includegraphics[width=0.8\linewidth]{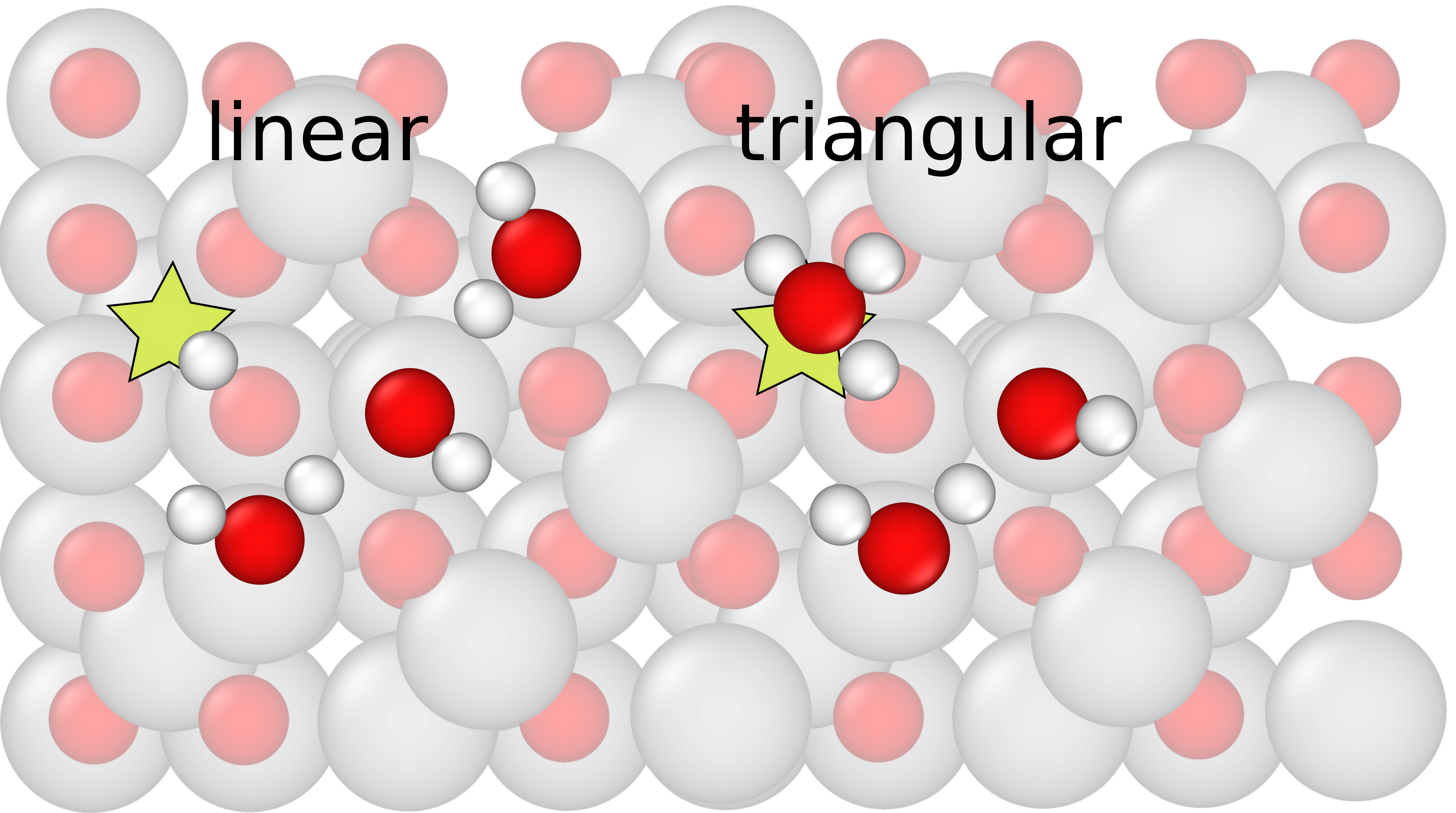}
    \caption{Linear and triangular water trimers on the SCV of magnetite. In both the configurations, a water molecule dissociates, donating an hydrogen to the surface oxygen near the dissociation site (indicated by the star). }
    \label{fig:trimers}
\end{figure}

Experiments including atomic force microscopy and scanning tunneling microscopy in combination with DFT calculations have shown that partially dissociated water clusters of various sizes form at the SCV~\citep{meier2018water}. One open question remains about the structure of the water trimer. In experiments, only linear trimers were observed. However, DFT calculations suggested that such linear trimers have the same adsorption energy as triangular trimers. To address this issue, we have simulated the SCV in contact with three water molecules in one unit cell and in a replicated two times two unit cell, both at constant 300 K. The starting water configurations are the two types of trimers on the two exposed SCVs (as shown in Fig.~\ref{fig:trimers}). The systems ran 54 ns and 33.5 ns respectively for the unit cell and the replicated one.

These simulations reveal that the linear and triangular trimers exhibit distinct dynamics. To quantify this difference we classified the water structures into clusters of different sizes at each timestep, utilizing a sparse matrix to encode pair distances with a cutoff of 3.5 ~\AA. 
By identifying clusters of three water molecules at the surface and distinguishing between the linear and triangular geometry, we calculated the probability of each trimer conformation. This probability was determined as the ratio of time spent in a particular conformation and the total time spent in any trimer form. While the results for the two independent surfaces agreed up to 0.01, for the two systems are very different from each other: the probability of being in a linear trimer is 0.72 for the three water in one unit cell and 0.97 for the replicated system with three water. This discrepancy indicates a large finite size effect due the interaction of the water cluster with its periodic image. Note that the water coverage of the two system also is also different (3 $\rm H_2O/u.c $ and 0.75 $\rm H_2O/u.c$ ). If we consider isolated trimers at the surface then we can write that: 
\begin{equation}
    \frac{P({\rm linear})}{P({\rm triangular})} \sim 32
\end{equation}
that is the linear trimer is more then thirty times more likely to occur than the triangular trimer at 300K. This effect is likely entropic in origin because the potential energy of the two trimer configurations is essentially the same.

\subparagraph{Water ground states}

\begin{figure}[]
    \centering
    \includegraphics[width=\linewidth]{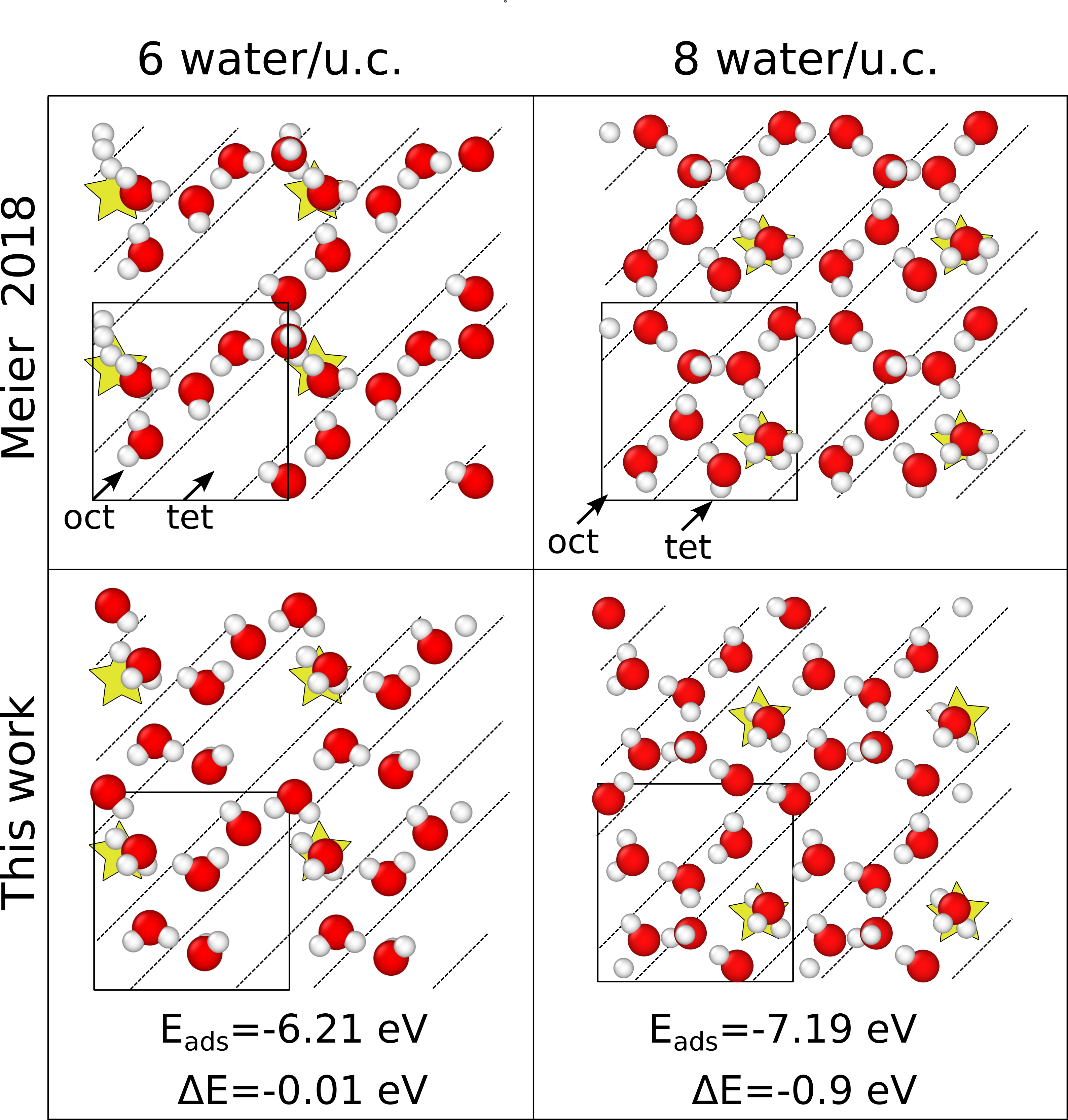}
    \caption{Minimum energy configurations of water on the SCV at low coverage. The configurations on the left involve six water molecule per unit cell while on the right eight. The minima in the top row are from~\citep{meier2018water} the ones in the bottom row 
    are new stable configurations discovered using the NNP. The SCV is not shown for a better visualization of the water structure but the iron rows are indicated by the dashed lines and the dissociation site by the stars. The number of dissociated hydrogens in the minima found in are two while the new ground states (with $\Delta E < 0$) have only one dissociated hydrogen. }
    \label{fig:stable_waters}
\end{figure}

Next, we investigated the stable structures of the water on the SCV towards the monolayer. Previous configurations were found ~\citep{meier2018water} to preferentially arrange as shown at the top of Fig.~\ref{fig:stable_waters} for 6 and 8 water molecules per surface unit cell, respectively. We ran MD simulations at 200~K and 300~K for systems comprising one unit cell with 6 and 8 $\rm H_2O$ starting from both random configurations and the stable structure reported in Ref.~\citep{meier2018water}. During the MD runs, the systems with random initial structures relaxed and one water molecule dissociated while the systems initiated with the stable configurations eventually drifted away from the initial state. After gathering a total of 1522 ns from five independent trajectories for each of the two temperatures and coverages, we picked configurations equally spaced in time. Since two surfaces are involved in the simulation (one on either side of the magnetite slab), two independent water arrangements were produced for each configuration. Then, the obtained structures were relaxed by minimizing the energy using the NNP. This procedure was computationally very efficient with the NNP, allowing a quick initial energy comparison of a large number of structures (on the order of tens of thousands). Subsequently, the 100 lowest energy structures were minimized via DFT. All the candidates which showed lower energy than the already established configurations were selected and their water structures were sliced out and placed on top of the exact same magnetite slab in order to isolate the water contribution. Finally, these setups were further minimized via DFT. Following this procedure, we found new configurations, shown in the bottom row in Fig.~\ref{fig:stable_waters} for 6 $\rm H_2O$/u.c. (left) and for 8 $\rm H_2O$/u.c. (right), with comparable and lower energies ($\Delta E = -0.01$~eV and $\Delta E = -0.9$~eV respectively) to the previously reported ones (top row). The new structures are very different and while the previous minima have two dissociations and two water ions per unit cell, the new ground states have only one dissociated water molecule and one water ion per unit cell. The adsorption energy is defined as
\begin{equation*}
    E_{\text{ads}}(n)= E_{\text{tot}} - E_{\text{slab}}- n E_{H_2O},
\end{equation*}
where $n$ is the number of water molecules in the system, and it contains the bond energy of the water-magnetite and water-water interactions. The adsorption energy was computed for the new minima:
\begin{equation*}
    E_{\text{ads}}(6)= -6.21 \text{ eV}, \quad E_{\text{ads}}(8)=-7.19 \text{ eV}
\end{equation*}

\subparagraph{Dissociation mechanism}

To reveal details of the mechanism of the water dissociation, we have simulated systems with 1 $\rm H_2O$/u.c. and 4 $\rm H_2O$/u.c. randomly distributed in a 3 by 3 replicated cell for roughly 100 ns and with 6 and 8 $\rm H_2O$/u.c. in a 6 by 6 replicated cell (to avoid finite size effects) for 50~ns. All simulations were carried out at a temperature of 300 K.


The first requirement for a dissocation to occur is that a water molecule positions itself on top of the dissociation site (see the 'no dissociation' panel instert in Fig. \ref{fig:diss}). At this point, the dipole vector of the water molecule points towards the surface but the molecule  will not yet dissociate. For that to happen, another water molecule must adsorb on a neighboring octahedral iron. Then, the water molecule on top of the vacancy dissociates, taking one hydrogen atom from its water neighbor. As a result, a hydronium forms on top of the dissociation site, whereas a hydroxide stays on top of a neighbor octahedral iron. Finally, the hydronium gives the extra hydrogen to an oxygen from the magnetite surface within the dissociation site environment (for simplicity we call this hydrogen atom the {\em dissociated hydrogen} in the following). This behavior is observed for all water coverages, but what happens next depends on the coverage. For coverages below 4 $\rm H_2O$/u.c., dissociation events lead water to stabilize in clusters while systems with higher coverages allow the hydroxide left from the dissociation to move along the octahedral iron rows, which are fully (or almost fully) occupied by water molecules. The dissociation mechanism has important consequences on the diffusion of water on the SCV.

\subparagraph{Water mean square displacement and diffusion}

\begin{figure}[]
    \centering
    \includegraphics[width=\linewidth]{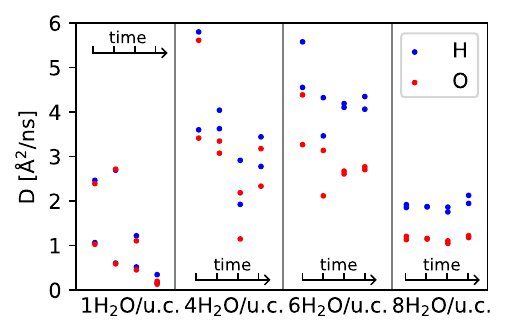}
    \caption{Diffusion coefficient $D$ of oxygens (red) and hydrogens (blue) for a coverage of 1,  4 , 6, and 8 $\rm H_2O$/u.c.. The diffusion coefficient is obtained from segments of the respective trajectories after an equilibration period. The results are shown for both the two independent interfaces present in the simulation setup.}
    \label{fig:diffusion}
\end{figure}

\begin{figure}[]
    \centering
    \includegraphics[width=\linewidth]{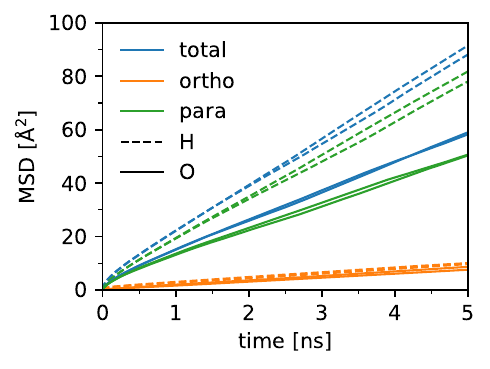}
    \caption{Mean square displacement MSD  (total, parallel and orthogonal to the iron rows) of water oxygens (full line) and hydrogens (dashed lines) at the magnetite SCV interface with 6 water molecules per unit cell. The results are shown for the two surfaces of the slab. }
    \label{fig:msd}
\end{figure}

\begin{figure*}[t!]
    \centering
    \includegraphics[width=\linewidth]{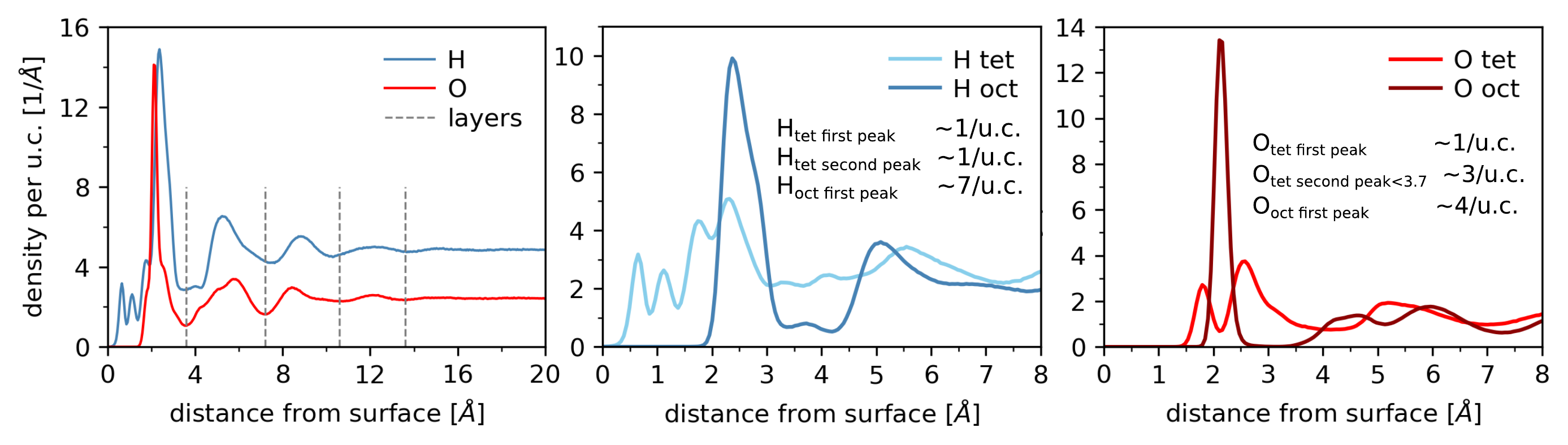}
    \caption{Density profiles of water at the magnetite surface as a function of the distance from the interface (the reference is the average $z$-position of the Fe and O atoms of the magnetite surface). The plot on the left shows the density of H and O separately. At very short distances from the interface, two small peaks are observed for the hydrogen density, arising from the dissociation of water molecules. Slightly above 2 \AA, the first oxygen peak appears along with a high hydrogen peak. The dashed grey lines indicate the minima in the oxygens density and subdivide the water in layers. The plots in the middle and on the right show the density profile of H and O respectively, distinguished by their location on top of octahedral or tetrahedral iron rows. For the oxygen, the first layer (up to 3.5 \AA) is characterized by a high, sharp, and well-separated octahedral peak flanked by two tetrahedral peaks. In the second layer (from 3.5~{\AA} to 7.2 \AA) the tetrahedral and octahedral regions are still non-equivalent, as the total peak decomposes into two octahedral peaks and one tetrahedral peak.}
    \label{fig:density}
\end{figure*}

 From the same batch of trajectories of the previous paragraph, the average mean square displacements (MSD) of the oxygens and hydrogens of the water molecules were computed. The MSDs were determined separately for the molecules on the two surfaces of the magnetite slab. After an equilibration time of 20~ns (10~ns) for 1 and 4 $\rm H_2O$/u.c. (6 and 8 $\rm H_2O$/u.c.), we have extracted four values of the diffusion coefficients from linear fits to the MSDs. In order to determine the accuracy of the diffusion constant, we have divided each trajectory in four equal parts and have carried out the analysis for each of them. The results are shown in Fig.~\ref{fig:diffusion}.  The diffusion coefficients of hydrogen and oxygens of water in system with 1 $\rm H_2O$/u.c. decreases with time due to the  dissociation of water molecules: once water clusters stabilize by dissociating the hydrogen, water dimers and trimers are pinned near the dissociation site. As the recombination event is extremely rare at 300\,K, these water molecules are essentially immobile for the entire simulation. At this stage, contributions to the mean square displacement come from the rest of the water molecules that roam detaching from one water cluster and attaching to another. The more water molecules a cluster has, the higher is the probability that a water at its border escapes. The pinned atoms are not the same during the simulations as waters can exchange. Therefore, for water agglomerates, the kinetics of water molecules consists of a sequence of hops, in which particles are trapped for a while until some cooperative motion unlocks a path in the vicinity and allows a few molecules to jump away. This mechanicm explains why the diffusion coefficient drops to a low value for 1 $\rm H_2O$/u.c systems. In the system with 4 $\rm H_2O$/u.c., the diffusion coefficient is very different for the two surfaces even after 100 ns, implying that the diffusion coefficient is affected by considerable statistical uncertainties in this case. For 6 and 8 $\rm H_2O$/u.c. the diffusion coefficient converges for the two surfaces over time. Averaging the values of the diffusion coefficient of the last two runs (those equilibrated) of 6 $\rm H_2O$/u.c. and all the 8 $\rm H_2O$/u.c. runs we obtained the following diffusion coefficients given in units of in [Å$^2$/ns]: 
 \begin{equation}
 \begin{split}
       &D_{\rm H}^{6\rm H_2O/u.c }=4.2\pm 0.1 \quad \quad D_{\rm O}^{6\rm H_2O/u.c }=2.7\pm 0.05 \\&
       D_{\rm H}^{8\rm H_2O/u.c }=1.9 \pm 0.1 \quad \quad D_{\rm O}^{8\rm H_2O/u.c }=1.2\pm 0.05
 \end{split}
 \end{equation}
 
Notably, the diffusivity of the hydrogen atoms is larger than that of the oxygen atoms. The reason for this is again related to dissociation. The presence of OH$^-$ ions in the water layer stimulates hydrogen exchanges, resulting in an overall higher mobility for the hydrogen atoms. Hence, hydrogen atoms have two ways to diffuse: one as a part of a water molecule and the other one as hydrogen transfer, (H$_2$O+OH$^-$ $\rightarrow$ OH$^-$+H$_2$O). 
 
We have also computed the directional MSDs parallel and orthogonal to the iron rows for the systems with 6 and 8 $\rm H_2O$/u.c as shown in Fig.~\ref{fig:msd}. The ratio of parallel and perpendicular diffusion coefficients of the oxygen is:
 \begin{equation}
       \frac{ D_{\parallel}^{6\rm H_2O/u.c }}{D_{\perp}^{6\rm H_2O/u.c }}\sim 6 \quad  \quad     \frac{ D_{\parallel}^{8\rm H_2O/u.c }}{ D_{\perp}^{8\rm H_2O/u.c }} \sim 7.5 
 \end{equation}
Hence, water molecules diffuse mainly along the iron rows.




\subsection{The magnetite-water interface}

Using he committee of NNPs, we simulated the magnetite SCV in contact with a slab of liquid water. The setup was a four times four replicated magnetite unit cell with a column of 2176 water molecules which was roughly 56~{\AA} high. The simulation was conducted for 29 ns, during which no disagreements in the committee force predictions were encountered. The temperature was constant at 300~K and the pressure was set to 0 bar. During the initial equilibration phase lasting about 10 ns, the water organizes and dissociates at the surface, eventually reaching a stationary regime. The analysis presented below focuses on the remaining 19 ns of the trajectory. We examine the statistical distribution of configurations of water at the surface, averaging over time and the two surfaces of the magnetite slab.

In particular, we computed the density profile of oxygen and hydrogen atoms in the liquid water phase above the SCV (see Sec.~\ref{sec:nnp simulations deteails} for details). As can be seen in Fig.~\ref{fig:density}, the water molecules are organized differently than in isotropic bulk water up to 15~\AA~from the interface. Focusing on the oxygen distribution, a distinct layering of water at the surface can be observed: four peaks and minima define four hydration layers located at a height 3.6, 7.2, 10.6, and 13.8~\AA. We find that the first two oxygen peaks consist of multiple sub-peaks, indicating the presence of a finer layering structure.

In order to refine the density profile, we distinguish between oxygens and hydrogens at octahedral and tetrahedral sites (details in Sec.~\ref{sec:nnp simulations deteails}). The corresponding density profiles, shown in Fig.~\ref{fig:density}, yield a more detailed understanding of the structure of the first water layer. 


\begin{figure}[h]
    \centering
    \includegraphics[width=\linewidth]{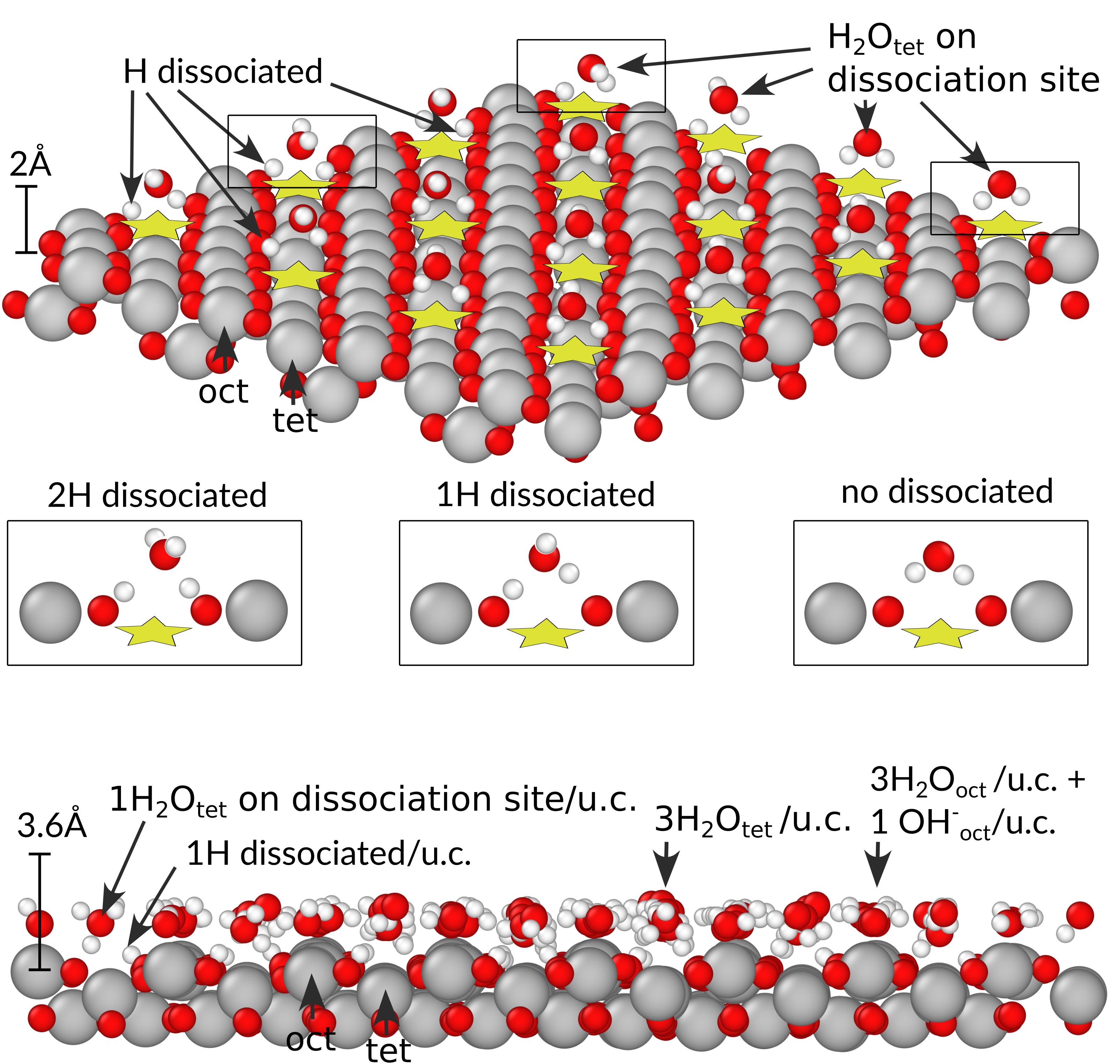}
    \caption{ Two cuts of the simulation at 2~{\AA} and 3.6~{\AA} (the covalent OH-bonds are not broken to preserve the water molecules) from the interface are shown in tilted view (top) and side view (bottom) respectively. The cut at 2~{\AA} height above the surface (top) includes all the dissociated hydrogens and all the tetrahedral water molecules on top of the dissociation site (stars). There are three possibilities (shown in the insert panels): either none, one or two dissociated hydrogens. The cut at 3.6~{\AA}  (bottom) is the complete first monolayer which includes the dissociated hydrogens, the water on top of the dissociation site and also 4 $\rm H_2O$/u.c. with one missing H (the one dissociated) and other 3 $\rm H_2O$/u.c. above the tetrahedral irons.}
    \label{fig:diss}
\end{figure}

The first tetrahedral hydrogen peak in the leftmost plot in Fig. \ref{fig:density} arises from dissociated hydrogens, which bend from their surface oxygen towards the dissociation site. The second peak is slightly wider but smaller and it is attributable to hydrogens that are less constrained in their movements and are involved in a hydrogen bond with the surface. The water molecules on top of the dissociation sites donate these H-bonds to the surface oxygens. If there are no dissociated hydrogen there are two H-bonds (insert right panel of Fig. \ref{fig:diss}), if there is only one dissociated hydrogen at the surface, there is one H-bond opposed to the dissociated OH (insert middle panel) if the dissociation site already presents two dissociated H no hydrogen bond is formed (insert left panel). By integrating the first two peaks, we expect the number of dissociations per unit cell to be 0.94 while the number of H-bonded hydrogen accounts for 0.93 atoms. The oxygens of the water molecules on the dissociation site produce the first well defined tetrahedral peak in the right hand plot in Fig. \ref{fig:density} that yields one atom if integrated up to 2.2~\AA~from the surface. This indicates the stability of the configuration with one dissociation sketched in the middle inserted panel in Fig.~\ref{fig:diss} and we can infer that it is far more probable (see also the whole snapshot of the simulation in Fig. \ref{fig:diss}).

The next peak to emerge corresponds to oxygens located along the octahedral iron rows. The sharpness of the peak, which integrates to four atoms, is a measure of the interaction of the water oxygen's with the magnetite irons. One water oxygen bonds to one octahedral iron, fully covering the row. The corresponding hydrogens also produce a sharp peak with a shoulder that shifts it to a larger distance. This implies that the OH bonds of the water molecules on the octahedral sites are partially horizontal on the plane and partially pointing away from the surface. The average number of hydrogens in the octahedral region of the first layer is 7, as one hydrogen is dissociated at the surface and the OH$^-$ is delocalized within the chain of water molecules adsorbed onto the octahedral iron rows. 

The oxygens above the tetrahedral irons complete the first layer. These are not anymore bonded to the surface as they produce a broad peak which starts only at 2.2~{\AA} and lacks a clear ending. Imposing an average of 3 tetrahedral oxygen atoms (one per tetrahedral iron), we infer that the first layer extends to about 3.7 ~\AA. The hydrogens needed in the first layer in the tetrahedral region to balance the charge and complete 4 water molecules are 7.3 (recalling that the tetrahedral water on the dissociation site has 0.93 H/u.c. under the second H peak), which is the value obtained when integrating up to the fifth tetrahedral peak (from 1.4 to 3.7 \AA). This indicates that the hydrogens of water molecules situated on top of tetrahedral irons are pointing toward the magnetite in an opposite fashion of the octahedral oxygens. The side view of the first layer is shown in the bottom panel of Fig. \ref{fig:diss}. The second layer is already less structured. Nevertheless, from the octahedral and tetrahedral splitting of the water density (Fig.~\ref{fig:density}), we can roughly estimate the average number of molecules in this layer. The two octahedral peaks account for 4.2 O/u.c. (from 3.5~{\AA} to 7.2 \AA), and the tetrahedral peak accounts for 4.1 O/u.c. (from 3.6~{\AA} to 7 \AA), resulting in a total of 8.3 O/u.c. in the second layer.

\section{Discussion and conclusion}
\label{sec:discussion}

In this work we have developed an NNP of Behler-Parrinello architecture for the magnetite/water interface. We have produced a first principles based dataset for SCV terminated magnetite in contact with water molecules, which we cleaned of ``corrupted'' data encountered due to the instability of DFT calculations (see secion \ref{sec:clean}). As the manual inspection was not sufficient to spot every error, we leverage the committee of NNPs to remove unsuitable data from the dataset. Although this procedure may seem arbitrary it proved very effective. The production of a cleaned dataset and the training of the NNP enabled stable simulations with first principles accuracy of the SCV in contact with water. The final NNP is able to handle all the water coverages, from bare surfaces to fully covered slabs, across a wide temperature range up to 400~K (for simpler system as one and two water per unit cell one can also safely run at 500~K). The NNP also describes the dissociation of water molecules occurring at the surface.
Employing the new committee of NNPs we simulated the SCV in contact with water agglomerates and liquid water. From the analysis of the trajectory of the 3 $\rm H_2O$/u.c. system, two energetically equivalent water trimers could be distinguished dynamically. The linear trimer, which is the only trimer observed in experiments~\citep{meier2018water}, is more than twice as likely than the triangular trimer. Our NNP-based simulations detected new ground states of water on the SCV with 6 and 8 $\rm H_2O$/u.c.. These structures need in the future to be verified with XPS or AFM experiments. The exploration of the energy minima for higher level of coverage is also an intriguing continuation. As the experimental results indicate that for higher coverages the water ground state could break the one unit cell periodicity~\citep{meier2018water}, it would be interesting to compare the newly found minima with minima obtained from an NNP-based simulation of a two times two replicated unit cell. Our NNP-based simulations also provided useful insights into the dissociation mechanisms and we obtained qualitative and quantitative information on the diffusion of water molecules on the SCV. The extensive MD simulations of the  magnetite-water system allowed to gather enough statistics to determine a detailed density profile. Our results indicate that water organizes at the interface in four different layers visible in the oxygen's density profile and recovers bulk-like behaviour only after 15 ~\AA. Incorporating geometrical information of the SCV, in particular dividing the water environment depending on its location above tetrahedral or octahedral iron rows, yields a comprehensive understanding of the structure of water at this interface. Important properties of the magnetite/water interface, such as its vibrational spectroscopy and the dissociation kinetics of water molecules are left for future study. 

 A limitation of the developed NNP lies in its restriction to water on the SCV model. This weakness hinders the exploration of the magnetite exposure to water in other directions. Nevertheless, we are confident that, by augmenting the training set with structures featuring different terminations, our model could be generalized to encompass alternative surfaces (like the (111) termination \citep{li_adsorption_2016, zaki_water_2018, santos-carballal_dft_2014}), potentially unveiling new stable surface structures and reconstructions.

Another subject for a future study would be the comparison of our results with more recent ML models trained on the same dataset, for instance using equivariant ML potentials~\citep{dusson_atomic_2022, batatia_foundation_2024, musaelian_learning_2023}.

\begin{acknowledgments}
This research was funded by the Austrian Science Fund (FWF) 10.55776/F81. For Open Access purposes, the authors have applied a CC BY public copyright license to any author accepted manuscript version arising from this submission. The computational results presented were achieved using the Vienna Scientific Cluster (VSC).
\end{acknowledgments}

\bibliography{newbib.bib}

\end{document}